\documentclass[3p,times,twocolumn]{elsarticle}

\usepackage{ecrc}

\volume{00}

\firstpage{1}

\journalname{Nuclear Physics B Proceedings Supplement}

\runauth{S.~Oh}

\jid{nppp}

\jnltitlelogo{Nuclear Physics B Proceedings Supplement}

\usepackage{amssymb}
\usepackage{lineno}

\usepackage[figuresright]{rotating}
\graphicspath{{./img/}}
\begin{document}

\begin{frontmatter}



\dochead{}

\title{Forward-central two-particle correlations in p--Pb collisions \\ 
with ALICE at the LHC}


\author{Saehanseul Oh for the ALICE Collaboration}
\address{Yale University, New Haven, CT 06520, USA}

\begin{abstract}
We present the results of measurements of two-particle angular correlations between trigger particles reconstructed in the ALICE Forward Muon Spectrometer ($-4.0 < \eta < -2.5$) and associated particles reconstructed in the central barrel detectors ($|\eta| < 1.0$) in p--Pb collisions at a nucleon-nucleon center-of-mass energy of 5.02 TeV.  
At low transverse momentum ($p_{\mathrm{T}}$), the reconstructed trigger particles mainly originate from weak decays of primary pions and kaons, whereas at high $p_{\mathrm{T}}$, they originate mainly from the decay of heavy-flavor particles. 
The ridge structure elongated in the $\eta$-direction, discovered recently in p--Pb collisions, is found to persist to the pseudorapidity ranges studied here. 
The second-order Fourier coefficients of muons are extracted after subtracting the correlations obtained in low-multiplicity events from those in high-multiplicity events. 
The ratio of coefficients in the Pb-going and p-going directions is presented as a function of $p_{\mathrm{T}}$, and the coefficients are observed to have a similar $p_{\mathrm{T}}$ dependence in both directions with the Pb-going coefficients larger by 16$\pm$6\%. 
The results are compared with calculations from a multi-phase transport model.
\end{abstract}

\end{frontmatter}

% \linenumbers
%\begin{linenumbers}

%% Introduction
\section{Introduction}
\label{introduction}
Two-particle angular correlations in $\Delta\varphi$ and $\Delta\eta$, where $\Delta\varphi$ and $\Delta\eta$ denote differences in the azimuthal angle ($\varphi$) and the pseudorapidity ($\eta$) between trigger and associated particles, have been widely used to explore particle production mechanisms in the field of relativistic heavy ion physics. 
In nucleus-nucleus collisions, ridge-like structures on the near (at $\Delta\varphi \approx 0$) and away side (at $\Delta\varphi \approx \pi$) along the $\Delta\eta$ axis are generally considered to be essential evidence of collective behavior of the created medium \cite{Aamodt:2011by}. 
Recently, these ridge-like structures were also observed in high-multiplicity proton--proton \cite{Khachatryan:2010gv} and proton-lead (p--Pb) collisions \cite{Abelev:2012ola, CMS:2012qk, Aad:2012gla} at LHC energies. 
These observations have opened new debates on their origin. 
Various theoretical attempts have been made to explain these results, such as models based on hydrodynamic flow, color glass condensate, and multi-parton interactions. 

One way to further understand these ridge-like structures is to investigate their $\eta$-dependence in p--Pb collisions. 
The CMS Collaboration has reported preliminary results evaluating the azimuthal harmonic coefficients from particle correlations within $|\eta|<2$, and results indicate a mild $\eta$-dependence within their acceptance \cite{CMS-PAS-HIN-14-008}. 
In addition, predictions based on a 3+1 dimensional viscous hydrodynamical model and a multi-phase transport model (AMPT) at large $\eta$ ($2.5 < |\eta| < 4$) show a strong $\eta$-dependence in the harmonic coefficients \cite{Bozek:2015swa}. 

ALICE has measured angular correlations between trigger particles from the Forward Muon Spectrometer (FMS) and associated particles from the central barrel detectors, which cover $-4.0<\eta<-2.5$ and $|\eta|<1.0$, respectively, in p--Pb collisions at $\ensuremath{\sqrt{s_{\mathrm{NN}}}}=5.02$ TeV. 
The analysis procedures and results are described in the following sections, while more details can be found in Ref.~\cite{Collaboration:2015il}. 

%% Experimental setup and analysis
\section{Experimental setup and analysis}
The current analysis is based on the p--Pb collision data collected by the ALICE detector in 2013 with a 4~TeV proton beam and a 1.58~TeV/nucleon lead ion beam. 
Beams were provided in both configurations, denoted by p--Pb (Pb--p) for proton beam going in the direction of negative (positive) $\eta$, allowing to measure muons in the p- and Pb-going directions.  
Due to the asymmetric beam energies, the nucleon-nucleon center-of-mass system is shifted in rapidity by 0.465 in the direction of the proton beam.
In order to ensure consistency of the multiplicity classes in both beam configurations, only symmetric parts of the V0 detectors ($2.8 < \eta < 3.9$ in V0-A and $-3.7 < \eta < -2.7$ in V0-C) were used for the event classification.

The analysis followed a similar procedure as in a previous measurement~\cite{Abelev:2013wsa} to evaluate $v_{2}^{\mu}\{\mathrm{2PC, sub}\}$, which is the second-order Fourier coefficient of the muons extracted from two-particle correlations with the subtraction method. 
The associated yield per trigger particle was measured in a given event class using muon tracks as trigger particles and inclusive charged hadrons in the central barrel detectors as the associated particles. 
The yield from the highest multiplicity class~(0--20\%) was subtracted by the yield from the lowest multiplicity class~(60--100\%) to isolate the long range correlations, and fitted with 
\begin{equation}
a_0 + 2\,a_1 \cos (\Delta\varphi) + 2\,a_2 \cos (2\Delta\varphi) + 2\,a_3 \cos (3\Delta\varphi)\,. 
\end{equation}
The relative modulation, $V_{n\Delta}\{\mathrm{2PC, sub}\}$, is given by $\frac{a_n}{a_0 + b}$, where $b$ is the baseline of the low-multiplicity class~(60--100\%) 
estimated from the integral of the per-trigger yield around the minimum. 
$v_{2}^{\mu}\{\mathrm{2PC, sub}\}$ are then obtained as
\begin{eqnarray}
v_{2}^{\mu}\{\mathrm{2PC, sub}\} = V_{n\Delta}\{\mathrm{2PC, sub}\} / \sqrt{V_{n\Delta}^{\mathrm c}\{{\mathrm{2PC,sub}\}}},
\end{eqnarray}
where $V_{n\Delta}^{\mathrm c}\{{\mathrm{2PC,sub}\}}$ is measured by correlations among inclusive charged hadrons in the central barrel detectors. 

Muon tracks used as trigger particles in the analysis were selected according to the same selection criteria as in Ref.~\cite{Abelev:2012pi} in the range of $0.5 < p_{\mathrm{T}} < 4.0$ GeV/$c$. 
The front absorber of the FMS is designed to suppress the contribution of weak decays from light hadrons, and affects the parent particle composition accordingly. 
\begin{figure}[t]
\centering
\includegraphics[width=0.44\textwidth]{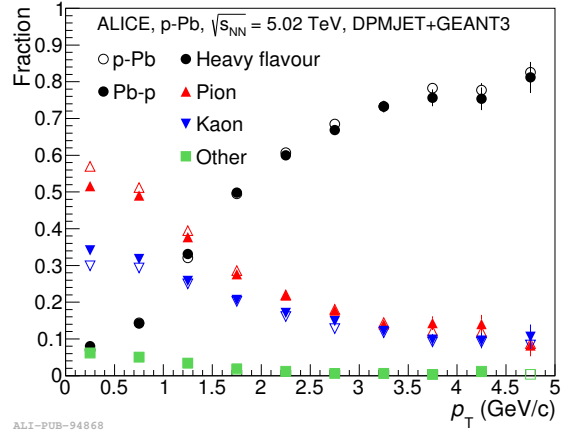}
\caption{\label{fig:mumothers}
Parent particle composition for reconstructed muon tracks in the p-going (p--Pb) and Pb-going (Pb--p) directions from detector response simulation of the ALICE Forward Muon Spectrometer with DPMJET Monte Carlo events. 
}
\end{figure}
Figure~\ref{fig:mumothers} shows the composition of parent particles of reconstructed muons in the p-going (indicated by p--Pb) and Pb-going (Pb--p) directions. 
This composition as a function of reconstructed $p_{\mathrm{T}}$ has been obtained with the DPMJET Monte Carlo (MC) event generator~\cite{Roesler:2000he} and GEANT3~\cite{geant3ref2} for the detector response simulation. 
While weak decays of pions and kaons dominate the low $p_{\mathrm{T}}$ region, their contributions are significantly reduced at $p_{\mathrm{T}} > 2$ GeV/$c$ and decays of heavy-flavor particles are dominant in this region. 
The parent particle compositions in the p-going and Pb-going directions are similar in this simulation. 

Both muon-track and muon-tracklet correlations were measured using central-barrel tracks and Silicon Pixel Detector (SPD) tracklets, repsectively, as associated particles. 
Central-barrel tracks, reconstructed based on the information from the Inner Tracking System (ITS) and the Time Projection Chamber (TPC), were selected in the same way as in Ref.~\cite{Abelev:2012ola} with the fiducial region $|\eta|<1$ and $0.5<p_{\mathrm{T}}<4.0$ GeV/$c$. 
SPD tracklets were reconstructed based on the primary vertex position and two hits on the SPD layers with $|\eta|<1$. 
Two hits were required to have an azimuthal angle difference less than 5 mrad.
This requirement suppresses fake and secondary track contributions and caused the mean $p_{\mathrm{T}}$ of selected tracklets to be about 0.75~GeV/$c$, which was estimated using the DPMJET MC. 

While $v_{2}^{\mu}\{\mathrm{2PC, sub}\}$ from muon-tracklet correlations was measured for both p-going and Pb-going directions, $v_{2}^{\mu}\{\mathrm{2PC, sub}\}$ from muon-track correlations was only calculated for the p-going direction due to low statistics. 
More details on ALICE detectors, such as the TPC, ITS, FMS and V0, can be found in Ref.~\cite{Aamodt:2008zz}. 

%% Results
\section{Results}
\label{Results}
\begin{figure}[t]
\centering
\includegraphics[width=0.44\textwidth]{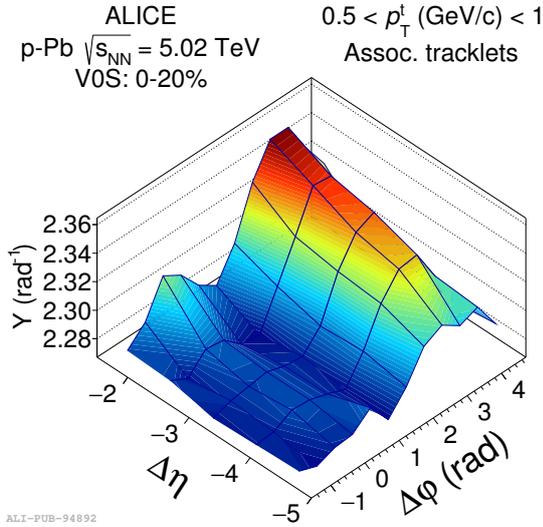}
\caption{\label{fig:yield}
An example of the $(\Delta\varphi, \Delta\eta)$ distribution of the associated yield per trigger particle in the highest multiplicity class using trigger particles from the FMS and associated particles from the central barrel detectors in p--Pb collisions.}
\end{figure}

An example of the associated yield per trigger particle is shown in Fig.~\ref{fig:yield}, for p-going muon tracks as trigger particles and SPD tracklets as associated particles in the 0-20\% event class. 
As the $\Delta\eta$ region is far away from the near-side jet-peak region around $(\Delta\varphi, \Delta\eta)=(0,0)$, no near-side jet contribution is observed. 
A near-side ridge structure is visible in this high multiplicity class. 

\begin{figure}[t]
\centering
\includegraphics[width=0.44\textwidth]{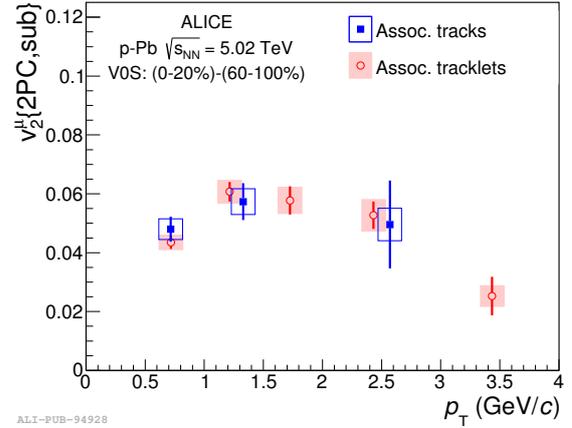}
\caption{\label{fig:v2comp}
$v_{2}^{\mu}\{\mathrm{2PC, sub}\}$ in the p-going direction, calculated for muons in $-4<\eta<-2.5$ from two-particle correlations with the subtraction method, extracted from muon-track and muon-tracklet correlations in p--Pb. Systematic and statistical errors are plotted as boxes and vertical error bars, respectively.
}
\end{figure}

$v_{2}^{\mu}\{\mathrm{2PC, sub}\}$ for p-going direction was measured with both tracks and tracklets as associated particles, separately, and is shown in Fig.~\ref{fig:v2comp}. 
Although the two different associated particle classes probe different $p_{\mathrm{T}}$ ranges, the $v_{2}^{\mu}\{\mathrm{2PC, sub}\}$ results agree with each other throughout the whole $p_{\mathrm{T}}$ ranges studied here. 
This agreement can be considered as evidence of the factorization between trigger and associate $v_{2}$. 

\begin{figure}[t]
\centering
\includegraphics[width=0.44\textwidth]{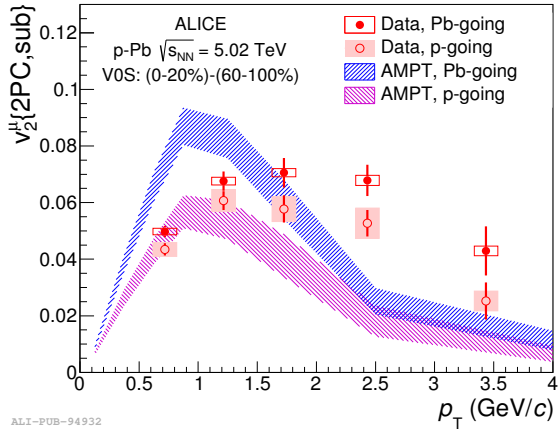}
\caption{\label{fig:v2final}
$v_{2}^{\mu}\{\mathrm{2PC, sub}\}$ from muon-tracklet correlations in the p-going and Pb-going directions for $-4<\eta<-2.5$ in p--Pb collisions at a center-of-mass energy of 5.02~TeV. Systematic and statistical errors are plotted as boxes and vertical error bars, respectively. Results from AMPT calculations are compared to the data. Only systematic errors are provided for the AMPT results.
}
\end{figure}

Using tracklets as associated particles, $v_{2}^{\mu}\{\mathrm{2PC, sub}\}$ for the p-going and Pb-going directions was measured and plotted in Fig.~\ref{fig:v2final}. 
The Pb-going $v_{2}^{\mu}\{\mathrm{2PC, sub}\}$ is observed to be larger than the p-going $v_{2}^{\mu}\{\mathrm{2PC, sub}\}$ throughout the measured $p_{\mathrm{T}}$ range, and both have a similar $p_{\mathrm{T}}$ dependence. 
$v_{2}^{\mu}\{\mathrm{2PC, sub}\}$ from AMPT model calculations are also provided in Fig.~\ref{fig:v2final}. 
It has to be noted that the predictions in Ref.\cite{Bozek:2015swa} based on a 3+1 dimensional viscous hydrodynamical model and the AMPT model cannot be directly compared with the data, although they cover the same $\eta$ range. 
Trigger particles used in the predictions are primary particles, and $v_{2}$ is extracted for primary particles while the presented $v_{2}^{\mu}\{\mathrm{2PC, sub}\}$ here is for decay particles after the front absorber of the ALICE FMS. 
This involves smearing effects in $\varphi$, and passing through the front absorber dramatically changes the parent particle composition for the muons relative to the particle composition of primary particles. 
Instead, the AMPT calculations in Fig.~\ref{fig:v2final} were prepared for the comparison with the data. 
Using the same parameters as in \cite{Bozek:2015swa}, generator level primary particles were simulated to decay into muons using the PYTHIA decayer \cite{Sjostrand:2006za}. 
The effects of the front absorber were accounted for by applying the relative efficiencies, which were evaluated for muons from pion and kaon decays relative to that for heavy flavor decay muons from a detector simulation of the ALICE FMS. 
The relative efficiencies can be found in Ref.~\cite{Collaboration:2015il}. 
Other details to calculate $v_{2}^{\mu}\{\mathrm{2PC, sub}\}$ in AMPT were the same as those in the data analysis, including event classification based on the number of charged particles in $2.8<\eta<3.9$ and $-3.7<\eta<-2.7$. 

The results from the AMPT calculation qualitatively agree with the data for $p_{\mathrm{T}} < 1.5$ GeV/$c$, the region dominated by weak decays of pions and kaons as shown in Fig.~\ref{fig:mumothers}. 
However, a qualitative discrepancy between data and AMPT calculations is observed for $p_{\mathrm{T}} > 2.0 $ GeV/$c$. 
As the contribution from heavy-flavor decays are the largest in this region and heavy-flavor $v_{2}$ in AMPT calculation is equal to 0, possible scenarios to explain this discrepancy can be either a finite value for the $v_{2}$ of the muons from heavy-flavor decays or large differences in particle composition and $v_{2}$ of each species between the AMPT generator and data. 

\begin{figure}[t]
\centering
\includegraphics[width=0.44\textwidth]{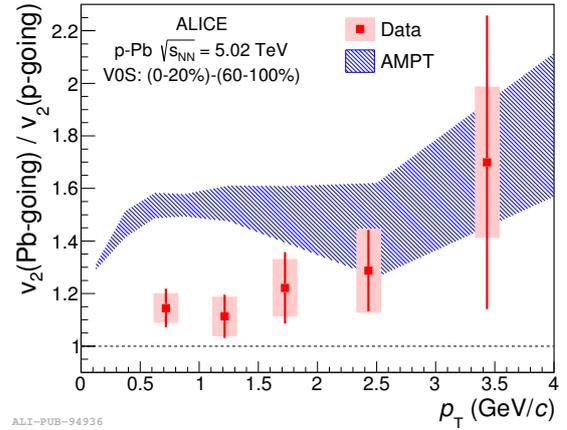}
\caption{\label{fig:v2ratio}
The ratio of $v_{2}^{\mu}\{\mathrm{2PC, sub}\}$ in the Pb-going direction and $v_{2}^{\mu}\{\mathrm{2PC, sub}\}$ in the p-going direction. The ratio from AMPT calculations is compared to the data. }
\end{figure}

To quantify the difference between $v_{2}^{\mu}\{\mathrm{2PC, sub}\}$ for the p-going direction and $v_{2}^{\mu}\{\mathrm{2PC, sub}\}$ for the Pb-going direction, the ratio of the two is calculated and shown in Fig.~\ref{fig:v2ratio}.
Considering the size of systematic and statistical errors, the ratio is observed to be almost independent of $p_{\mathrm{T}}$. 
A constant fit to the ratio is performed and returns the value, $1.16\pm0.06$, with a $\chi^{2}/$NDF=0.4.
Results from AMPT calculations show a similar $p_{\mathrm{T}}$ dependence.

All systematic uncertainties for Fig.~\ref{fig:v2comp}, \ref{fig:v2final}, and \ref{fig:v2ratio} are described in Ref.~\cite{Collaboration:2015il}. 

%% Summary
\section{Summary}
\label{summary}
The $v_{2}$ of muons reconstructed in the ALICE Forward Muon Spectrometer was measured from by two-particle correlations with subtraction method in p--Pb collisions at a nucleon-nucleon center-of-mass energy of 5.02 TeV. 
With Monte Carlo simulations, the composition of parent particles of reconstructed muons, largely affected by the front absorber, was determined   to vary depending on $p_{\mathrm{T}}$. 
$v_{2}^{\mu}\{\mathrm{2PC, sub}\}$ for the Pb-going direction was observed to be larger than $v_{2}^{\mu}\{\mathrm{2PC, sub}\}$ in the p-going direction by $16\pm6$\%, and both showed similar $p_{\mathrm{T}}$ dependence. 
Comparison with AMPT calculations revealed a disagreement with the data for $p_{\mathrm{T}}>2.0$ GeV/$c$, where muons from heavy-flavor decays dominate.

%% \end{linenumbers}
\appendix


%% References

\nocite{*}
\bibliographystyle{elsarticle-num}
\bibliography{Oh_S}






% \end{thebibliography}

\end{document}